\documentclass[a4paper]{article}

\usepackage{INTERSPEECH2020}
\usepackage{caption}
\usepackage{enumitem}
\usepackage[hidelinks]{hyperref}

\title{Generic Indic Text-to-speech Synthesisers with Rapid Adaptation in an End-to-end Framework}
\name{Anusha Prakash$^1$, Hema A Murthy$^2$}
\address{
  $^1$Dept of Electrical Engineering, Indian Institute of Technology Madras, India\\
  $^2$Dept of Computer Science \& Engineering, Indian Institute of Technology Madras, India}
\email{anushaprakash@smail.iitm.ac.in, hema@cse.iitm.ac.in}

\begin{document}

\maketitle
\begin{abstract}

Building text-to-speech (TTS) synthesisers for Indian languages is a difficult task owing to a large number of active languages. Indian languages can be classified into a finite set of families, prominent among them, Indo-Aryan and Dravidian. The proposed work exploits this property to build a generic TTS system using multiple languages from the same family in an end-to-end framework. Generic systems are quite robust as they are capable of capturing a variety of phonotactics across languages. These systems are then adapted to a new language in the same family using small amounts of adaptation data. Experiments indicate that good quality TTS systems can be built using only 7 minutes of adaptation data. An average degradation mean opinion score of 3.98 is obtained for the adapted TTSes. 

Extensive analysis of systematic interactions between languages in the generic TTSes is carried out. x-vectors are included as speaker embedding to synthesise text in a particular speaker's voice. An interesting observation is that the prosody of the target speaker's voice is preserved. These results are quite promising as they indicate the capability of generic TTSes to handle speaker and language switching seamlessly, along with the ease of adaptation to a new language.
 




\end{abstract}

\section{Introduction}

The current state-of-the-art text-to-speech (TTS) paradigm is end-to-end (E2E) speech synthesis \cite{tacotron, tacotron2, transformer_tts_2018}. It is an attractive platform to use as it alleviates the need for an alignment module and language-specific modules, such as grapheme-to-phoneme converters. Synthesisers are easily trained given only speech waveforms and corresponding text transcriptions. Combined with improved vocoders, such as WaveNet \cite{WaveNet2016} and WaveGlow \cite{waveglow}, the speech synthesis quality has become almost human-like. But end-to-end speech synthesisers require many hours of training data to generate good quality synthesis and are also computationally intensive. This is challenge in the Indian context, considering the wide linguistic diversity and the paucity of speech resources.



To overcome this problem, we propose to develop generic voices for Indian languages in an end-to-end framework and adapt them to languages with low amounts of speech data. Generic voices are built by pooling data across multiple languages, where each language is spoken by a native speaker. This kind of pooling results in a 3-fold perspective:
\begin{enumerate}
    \item Pooling together multiple monolingual data increases the amount of training data. This leads to more robust models, especially in the end-to-end context, resulting in better synthesis quality.
    \item Generic voices capture a superset of phonotactical variations, thus facilitating better adaptation to a new language.
    \item This kind of pooling is a multilingual and multi-speaker scenario. Extensive analysis is carried out to observe the systematic interactions between languages/speakers in the generic TTSes. We see that it is possible to synthesise speech in a target language, given the voice characteristics of a non-native speaker.
    
\end{enumerate}


In the current work, generic voices are trained by pooling data of languages belonging to the same language family-- Indo-Aryan and Dravidian. This is because these language families have very distinct characteristics. While Indo-Aryan languages are characterised by schwa deletion (deletion of inherent vowel ``a''), Dravidian languages are characterised by their agglutinative nature of scripts (multiple words are combined to form a single word, leading to longer utterances) \cite{AnushaIS16}. The issue of pooling is also compounded by the fact that these languages have different written scripts. To circumvent this, the multi-language character map (MLCM) \cite{Prakash2019_MLCM} and the common label set (CLS) \cite{ramani, arunTSD16} for Indian languages are adopted for character-based and phone-based text representations, respectively.

Generic voices are trained based on Tacotron2 architecture for text to mel-spectrogram conversion \cite{tacotron2}. WaveGlow is used as the vocoder to generate speech waveforms from the mel-spectrograms \cite{waveglow}. To enable speaker switching, x-vectors are used as speaker embedding in the TTS framework \cite{espnet}. Generic voices are adapted to different amounts of speech data. Systems are evaluated using mel-cepstral distortion (MCD) scores and subjective measures such as degradation mean opinion scores (DMOS). Results indicate that the generic voices are quite robust and facilitate speaker selection using x-vectors. With as little as 7 minutes of adaptation data, a good quality TTS system is built very quickly.
 
The rest of the paper is organised as follows. Section \ref{sec:lit_survey} reviews the related literature. A brief overview of the end-to-end speech synthesiser employing Tacotron2 and WaveGlow vocoder is given in Section \ref{sec:E2E}. The proposed techniques, combinations of experiments and analysis of results are presented in Section \ref{sec:proposed}. The work is concluded in Section \ref{sec:conclusion} with directions for future work.

\vspace{-0.2cm}
\section{Related work}
\label{sec:lit_survey}

In the literature, there are many approaches to training an average voice model (AVM) (also called generic voice), in a multilingual and multi-speaker scenario. In \cite{Latorre_MLMS_2006, Zen2012_Factorization}, AVMs are trained in the HMM framework, with \cite{Zen2012_Factorization} modeling speaker and language-specific characteristics using individual factor-specific transforms. In \cite{multiLingual_LSTM2016, Gutkin17, DemirsahinJG18}, an LSTM-RNN (long short-term memory recurrent neural network) based approach is explored for training the AVMs. A unified multilingual phoneme inventory is developed for several major South Asian languages in \cite{DemirsahinJG18}. In \cite{Himawav20_MLMS}, a feed-forward deep neural network (DNN) is used to train the acoustic model, along with language-specific duration models, and language and speaker embeddings. AVMs are also trained in end-to-end frameworks \cite{Bytes_Google2019, Zhang19_foreign, Baljekar2018}. \cite{Bytes_Google2019} explores a language-independent approach to training by converting text into a sequence of Unicode bytes. In \cite{Zhang19_foreign}, an adversarial loss term is included for training, to disentangle the speaker identity and linguistic content. This enables the AVM to speak fluently in a foreign language. In these studies, AVM achieves comparable or improved synthesis quality over monolingual systems. AVMs are also successfully adapted to new languages/speakers \cite{Latorre_MLMS_2006, Zen2012_Factorization, multiLingual_LSTM2016, Gutkin17}.

In most of the above mentioned literature, AVMs are trained with hundreds of hours of data. In contrast, the current work uses a maximum of 20 hours to train the AVM. In the context of training AVMs for Indian languages in an end-to-end framework, \cite{Baljekar2018} explores convolution attention based models with global embeddings such as speaker, language and gender embeddings. Preliminary cross-lingual experiments are carried out, but they are not extensively studied. The current work analyses interactions between Indian languages in average or generic voices with speaker embedding, and also adapt generic voices to minimal amounts of data in new languages. Good quality synthesised speech is obtained from the adapted TTSes, with the voice characteristics of the target speakers being preserved. 




\vspace{-0.2cm}
\section{End-to-end speech synthesis}
\label{sec:E2E}

The end-to-end speech synthesis approach consists of two stages: text to mel-spectrogram conversion and speech waveform generation by conditioning on mel-spectrogram. In the current work, Tacotron2 is used for the first stage, and WaveGlow is used as the vocoder. These modules along with the incorporation of speaker embedding are briefly described in this section.

\vspace{-0.2cm}
\subsection{Tacotron2}
\vspace{-0.1cm}
Tacotron2 learns the mapping between linguistic and acoustic features during training. The basic architecture is a recurrent neural network (RNN) based sequence-to-sequence model \cite{tacotron2}. It consists of an encoder and an attention-based decoder. The encoder encodes the character embeddings extracted from the text. The attention module then summarises this variable-length encoded representation as a fixed-length context vector and feeds it to the decoder at each step. The decoder then predicts mel-spectrograms corresponding to frame(s) at each decoder step. A location-sensitive attention is used in the decoder \cite{attention_NIPS2015}. Additionally, a guided attention loss is included during training \cite{guided_attention}. This forces the attention weights to be diagonal, thus mimicking the monotonic nature of alignment between linguistic and acoustic features.
\vspace{-0.2cm}
\subsubsection{TTS with speaker embedding}
\vspace{-0.1cm}
The vanilla Tacotron2 framework does not explicitly include any speaker-specific information. In the pooled generic voice scenario, we cannot control the speaker's voice during synthesis. Hence, incorporating speaker embedding facilitates speaker selection. x-vectors are used for this purpose \cite{xvector_ASR, xvector_SV2017}. 

x-vectors are DNN embedding of a neural network that is trained to discriminate between speakers. They are fixed-length vectors extracted from variable length utterances and determined using a time-delay neural network (TDNN). During the training phase of TTS, x-vectors are obtained for each utterance. These speaker embeddings are replicated and appended to each encoder state in the sequence-to-sequence network. The TTS synthesiser is then trained similar to the vanilla version \cite{espnet}.

During the testing phase, only the test transcription is available. Per speaker x-vector is determined by taking the mean of x-vectors corresponding to that speaker in the training data. A test sentence is then synthesised in any speaker's voice by appending that speaker's x-vector to all the encoder states.
\vspace{-0.2cm}
\subsection{WaveGlow}
\vspace{-0.1cm}
WaveGlow is a non auto-regressive generative vocoder that produces fast and high-fidelity speech output \cite{waveglow}. Samples are taken from a zero mean spherical Gaussian and transformed through a series of layers to span the desired distribution. WaveGlow models the distribution of audio samples conditioned on mel-spectrograms. It is implemented using a single network and trained on a single cost function that minimises the negative log-likelihood of the data.


\vspace{-0.2cm}
\section{Generic voices and adaptation}
\label{sec:proposed}
\subsection{Datasets}
\vspace{-0.1cm}
The datasets used in this work are part of the IndicTTS database \cite{ArunResources2016}. For generic voices, constituent monolingual data of 5 hours each is used. Only one male speaker per language is considered due to the unavailability of multiple speakers for each language. Totally 9 languages spanning 8 different scripts are used in the experiments. Bengali, Gujarati, Hindi, Odia and Rajasthani are Indo-Aryan languages; while Kannada, Malayalam, Tamil and Telugu are Dravidian languages. All these languages have their own script, except for Hindi and Rajasthani, which share the Devanagari script.


Utterances limited to 15 seconds in duration are chosen for training,
as attention in end-to-end systems is not learnt very effectively for long training utterances, . The text is in UTF-8 format and is cleaned up as end-to-end training is very sensitive to data.

The subsequent sub-sections describe the training of Indo-Aryan and Dravidian generic TTSes, the incorporation of speaker embedding in the network, and finally, experiments on adaptation. Systems are evaluated based on objective measures-- dynamically time warped (DTW) mel-cepstral distortion (MCD) scores \cite{MCD} (lower scores are better) and using subjective measures such as degradation mean opinion score (DMOS) (a score of 5 indicates human-like quality) \cite{dmos}. MCD gives a measure of the spectral accuracy, and DMOS gives a measure of the quality of speech. For objective evaluation of each system, 20 held-out sentences that are not present in the training data are chosen. Examples of synthesised utterances for each experiment are given at this link: \url{https://www.iitm.ac.in/donlab/preview/average-voice-2020/index.html}.

\vspace{-0.2cm}
\subsection{Generic voices}
\vspace{-0.1cm}
\label{sec:generic}

\begin{figure}[ht!]
 \centering
 \includegraphics[width = \linewidth]{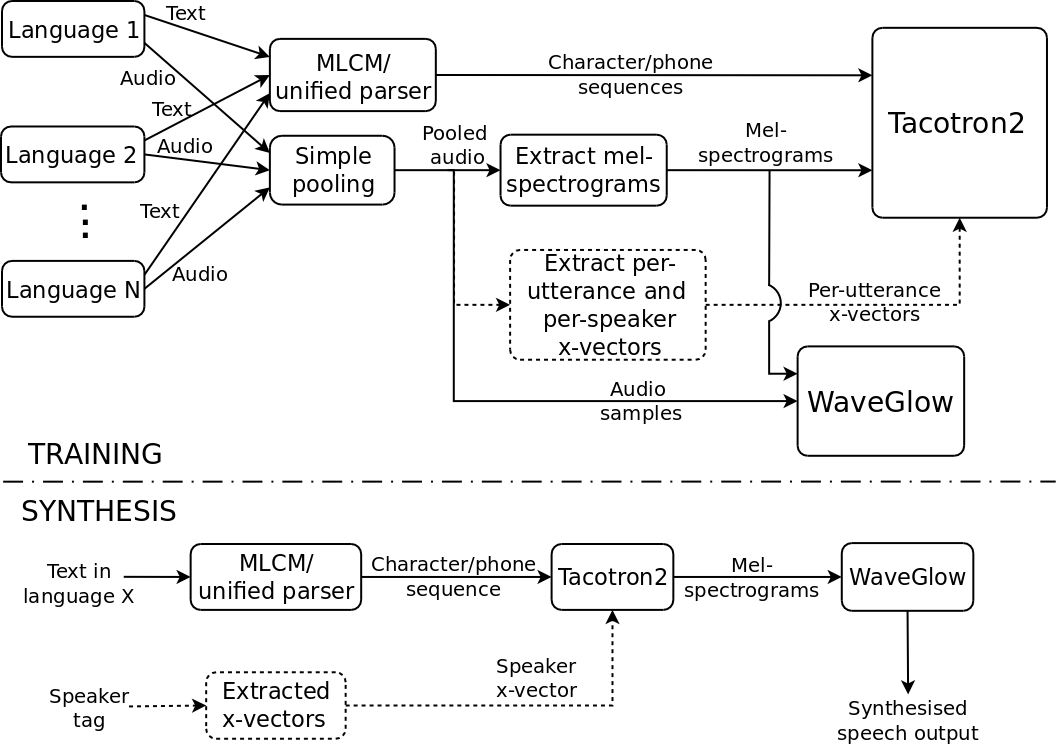}
 \caption{Training and synthesis phases of the generic TTS. Blocks with dashed lines are for system with speaker embedding.}
 \vspace{-0.4cm}
 \label{fig:system}
\end{figure}

Figure \ref{fig:system} illustrates the blocks involved in training a generic system and synthesis of a test sentence. Generic systems are trained by pooling multiple monolingual data, each recorded by different speakers. Experiments in \cite{Prakash2019_MLCM} show that when languages having different scripts are pooled, it leads to poor training. Hence, a common linguistic representation is required. For this, a multi-language character map (MLCM) is used \cite{Prakash2019_MLCM}. In MLCM, similar character representations across 13 Indian languages are mapped together. Training text is converted to character sequences using MLCM. Audio files are pooled together and mel-spectrograms are extracted from them. Tacotron2 learns the mapping between character sequences and mel-spectrogram frames. ESPnet's implementation of Tacotron2 with default hyperparameters is used in this work \cite{espnet}.

For Dravidian languages, instead of a character-based text representation, a phone-based representation is used. This is primarily to address the problem of Tamil having the same written representation for voiced and unvoiced stop consonants. MLCM does not have the provision to make this distinction as it is character-based. A unified parser for Indian languages \cite{arunTSD16} parses the text into a common label set representation \cite{ramani}, which is then converted to a phone-mapped representation as described in \cite{Prakash2019_MLCM}. 

To synthesise speech from mel-spectrograms generated by the Tacotron2 module, Nvidia's WaveGlow vocoder is used \cite{waveglow}. Since training a WaveGlow model from scratch requires a lot of computational time, the WaveGlow model is fine-tuned on a pre-trained LJ Speech model. For the generic voice, WaveGlow is trained on the pooled audio data. For all experiments, it was observed that WaveGlow produced synthesised speech characterised by line noise. This was eliminated using notch filters.

For the Indo-Aryan generic TTS, 5 hours each from four languages are pooled-- Bengali, Hindi, Odia and Rajasthani; the total training data is of 20 hours in duration. For the Dravidian generic voice, 5 hours each from three languages are pooled-- Kannada, Malayalam and Telugu, resulting in a total of 15 hours for training. Generic voices are compared to systems trained on monolingual data. MCD scores are calculated on monolingual sentences generated by generic TTSes and the corresponding monolingual TTS. Results given in Tables \ref{table:gen_Aryan} and \ref{table:gen_Drav} indicate that monolingual TTSes have lower MCD scores compared to generic TTSes in most cases. The slight drop in performance of generic TTSes may be attributed to the influence of language mixing.

Speaker embedding is not explicitly provided during synthesis; the target speaker's voice is internally chosen based on the script. This creates a conflict between Hindi and Rajasthani which share the same script. It was observed that few Rajasthani synthesised samples had the Hindi speaker's voice characteristics.

To evaluate the scalability of generic voices for new languages, Gujarati and Tamil sentences are synthesised by the Indo-Aryan and Dravidian generic TTSes, respectively. It is observed that synthesised samples of Gujarati have different speaker characteristics for different sentences, which is chosen internally based on context. On the other hand, Tamil synthesised samples have only Malayalam speaker's characteristics. Nevertheless, the synthesised utterances of these new languages are intelligible.  This indicates the ability of generic TTSes to synthesise unseen languages reasonably well, albeit with no control over speaker selection.


\vspace{-0.2cm}
\begin{table}[h!]
\centering
\footnotesize
\caption{Indo-Aryan: MCD scores of monolingual and generic TTSes (with and without speaker embedding)}
\label{table:gen_Aryan}
\vspace{-0.2cm}
\begin{tabular}{|l|r|r|r|r|}
\hline
Language    & Bengali & Hindi & Odia & Rajasthani \\ \hline
Monolingual &   7.305      &   7.789    &  7.501    &      7.049      \\ \hline
Generic      &   7.424      &   7.968    & 7.525     &      6.999      \\ \hline
Generic (x-vector)      &   7.437      &   7.664    &   7.418   &     6.606       \\ \hline
\end{tabular}
\end{table}

\vspace{-0.5cm}
\begin{table}[h!]
\centering
\footnotesize
\caption{Dravidian: MCD scores of monolingual and generic TTSes (with and without speaker embedding)}
\label{table:gen_Drav}
\vspace{-0.2cm}
\begin{tabular}{|l|r|r|r|}
\hline
Language    & Kannada & Malayalam & Telugu \\ \hline
Monolingual &     8.297    &      7.562     &    8.621    \\ \hline
Generic      &    8.269     &     7.723      &    9.282    \\ \hline
Generic (x-vector)      &    7.935     &    7.402       &    8.619    \\ \hline
\end{tabular}
\end{table}
\vspace{-0.4cm}

\vspace{-0.2cm}
\subsection{Generic voices with speaker-embedding}
\vspace{-0.1cm}
\label{sec:generic_xvec}

x-vectors are used as speaker embedding in the generic TTSes. The incorporation of x-vectors is shown with dashed modules in Figure \ref{fig:system}. 512-dimensional x-vectors are extracted from the audio files using a pre-trained x-vector \cite{xvector_ASR} provided by Kaldi. Experiments with speaker embedding in generic voices are presented here. Languages/speakers present in the pooled data are referred to as seen languages/speakers.

\textbf{\emph{(a) Seen language, seen speaker (without speaker switching)}}: Monolingual sentences are synthesised using the x-vector of the corresponding monolingual speaker. For example, Hindi sentences are synthesised with the Hindi speaker's x-vector. MCD scores of the synthesised samples are given in Tables \ref{table:gen_Aryan} and \ref{table:gen_Drav}. It is seen that generic voices with speaker embedding perform better with lower MCD scores compared to vanilla generic and monolingual TTSes. Additionally, the conflict between Hindi and Rajasthani speakers for the synthesis of a Devanagari text is resolved with explicit speaker embedding.

\textbf{\emph{(b) Seen language, seen speaker (with speaker switching)}}: Monolingual sentences are synthesised with a non-native speaker's x-vector. Hindi sentences are synthesised by the Indo-Aryan TTS with Bengali, Odia and Rajasthani speakers' x-vectors. Kannada sentences are synthesised by the Dravidian TTS with Malayalam and Telugu speakers' x-vectors. 

It is observed that the target speaker's voice and prosodic characteristics are preserved during the synthesis of a non-native text. Of the five text-speaker combinations, Hindi-Bengali and Kannada-Telugu were considered for evaluations. Listeners proficient in Bengali and Hindi (or Kannada and Telugu) were asked to listen to the Hindi (or Kannada) synthesis samples generated with Bengali (or Telugu) speaker's x-vector. The task was to select the nativity of the speaker -- Hindi/Bengali (or Kannada/Telugu). 11 listeners were asked to assess 10 sentences in each test. For the Hindi-Bengali test, listeners chose Bengali nativity with $81.82\%$ preference.  For the Kannada-Telugu test, listeners chose Telugu nativity with only $26.36\%$ preference, which was surprising. Listeners in this test indicated that it was hard to distinguish the nativity of the speaker. This may be because Kannada and Telugu speakers are closely associated with each other, compared to Hindi and Bengali. In a manner, the generic TTS system with speaker embedding does mimic the real-world scenario, as it captures systematic interactions between languages.

\textbf{\emph{(c) Seen language, unseen speaker}}: Hindi and Kannada sentences are generated from the Indo-Aryan TTS with Gujarati speaker's x-vector and from the Dravidian TTS with Tamil speaker's x-vector, respectively. x-vectors for the unseen languages, Gujarati and Tamil, are extracted from 5 hours of corresponding monolingual data. Although intelligible, synthesised speech does not have the target speaker's voice characteristics. The voice characteristics are somewhat averaged, dominated by seen speakers. In some cases, switching of speakers within an utterance is also observed.

\textbf{\emph{(d) Unseen language, unseen speaker}}: Experiments are similar to those in (c), with Hindi and Kannada test sentences replaced by Gujarati and Tamil sentences, respectively. In this case too, synthesised speech is intelligible, albeit with more of an averaged voice. 

\textbf{\emph{(e) Unseen language, seen speaker}}: Gujarati and Tamil sentences are generated from the Indo-Aryan TTS with Hindi speaker's x-vector and from the Dravidian TTS with Malayalam speaker's x-vector, respectively. Synthesised samples are intelligible and retain the voice characteristics of the target speaker. This is an improvement over vanilla generic systems which cannot control speaker characteristics for the synthesis of text in a new language (Section \ref{sec:generic}).
\vspace{-0.2cm}
\subsection{Adaptation}
\vspace{-0.1cm}
We see that even if a TTS synthesiser is trained with speaker embedding, it doesn't scale to unseen speakers. So the next step is to adapt the generic voice to adaptation data in/of the unseen language/speaker. The Indo-Aryan generic voice is adapted to Gujarati data and the Dravidian generic voice to Tamil data. It is to be noted that the Gujarati speaker used in the adaptation experiments is different from the one used for experiments in Section \ref{sec:generic_xvec}. Generic voices are adapted to 30 mins, 15 mins and 7 mins of adaptation data. The learnt parameters of the generic TTS are fine-tuned on the adaptation data. Experiments are performed by adapting generic voices, with and without speaker embedding. Baseline systems are TTSes trained on monolingual data from scratch, with the same amount of data used for adaptation.


Tables \ref{tab:guj_adapt} and \ref{tab:tam_adapt} show the MCD scores obtained for the adapted and baseline monolingual systems. It is clearly seen that the baseline systems perform the worst. These systems are trained very poorly due to the lack of adequate data. It is observed that both generic systems, with and without speaker embedding, when adapted to a new language/speaker, are able to generate utterances in the adapted speaker's voice. Of all systems, adaptation using the generic TTS system with x-vector performs the best. Comparing results across different amounts of adaptation data, the system trained using 7 minutes of data also performs well, given the minimal amount of data.

DMOS tests are conducted to assess the quality of synthesised speech generated by Gujarati and Tamil systems. These systems are adapted from the generic TTS with speaker embedding using only 7 minutes of data. 8  native Gujarati listeners and 19 native Tamil listeners evalauted 10 synthesised and 5 natural utterances in each test. Speaker similarity tests are also conducted, similar to DMOS, in which listeners rate utterances based on similarity to a reference speaker. Results in Table \ref{tab:dmos_adapted} show that good quality TTSes can be trained with very little data, while preserving the voice characteristics of the adapted speaker. This is also rapid adaptation, as these two systems were trained individually in about 30-45 minutes on a single GeForce GTX 1080 card.
\vspace{-0.1cm}
\begin{table}[h!]
\centering
\footnotesize
\caption{MCD scores of Gujarati TTSes trained using different amounts of adaptation data}
\label{tab:guj_adapt}
\vspace{-0.2cm}
\begin{tabular}{|l|r|r|r|}
\hline
Amount of Data                                                           & \multicolumn{1}{c|}{30 mins} & \multicolumn{1}{c|}{15 mins} & \multicolumn{1}{c|}{7 mins} \\ \hline
Monolingual                                                              & 11.240                       & 10.698                       & 10.888                      \\ \hline
Generic TTS Adapted                                                      & 7.592                        & 7.584                        & 7.567                       \\ \hline
\begin{tabular}[c]{@{}l@{}}Generic TTS Adapted\\ (x-vector)\end{tabular} & 7.486                        & 7.505                        & 7.538                       \\ \hline
\end{tabular}
\end{table}
\vspace{-0.5cm}
\begin{table}[h!]\
\centering
\footnotesize
\caption{MCD scores of Tamil TTSes trained using different amounts of adaptation data}
\label{tab:tam_adapt}
\vspace{-0.2cm}
\begin{tabular}{|l|r|r|r|}
\hline
Amount of Data                                                           & \multicolumn{1}{c|}{30 mins} & \multicolumn{1}{c|}{15 mins} & \multicolumn{1}{c|}{7 mins} \\ \hline
Monolingual                                                              & 10.534                       & 10.02                        & 10.471                      \\ \hline
Generic TTS Adapted                                                      & 8.301                        & 8.276                        & 8.503                       \\ \hline
\begin{tabular}[c]{@{}l@{}}Generic TTS Adapted\\ (x-vector)\end{tabular} & 8.231                        & 8.451                        & 8.494                       \\ \hline
\end{tabular}
\end{table}
\vspace{-0.5cm}

\begin{table}[h!]
\centering
\footnotesize
\caption{Results of adapted TTSes trained from generic TTSes with speaker embedding and 7 minutes of adaptation data}
\label{tab:dmos_adapted}
\vspace{-0.2cm}
\begin{tabular}{|l|r|r|}
\hline
TTS  & Gujarati & Tamil \\ \hline
DMOS &    4.41      & 3.54  \\ \hline
Speaker similarity &    3.95      & 3.2  \\ \hline
\end{tabular}
\end{table}
\vspace{-0.3cm}

\vspace{-0.2cm}
\subsection{Discussion}
\vspace{-0.1cm}
Good quality TTSes are trained by pooling data across languages and speakers. Pooling is possible in the Tacotron2 framework using MLCM and CLS to address the problem of different scripts. It is seen that these systems can scale to an unseen language, but speaker selection is not possible. With x-vectors, speaker selection is possible, but the system still doesn't scale to an unseen speaker. This is resolved by adapting the generic voice to a minimal amount of unseen speaker (or language) data. A further step would be to see how much of adaptation data can be reduced to train an adequate quality TTS.

\vspace{-0.2cm}
\section{Conclusion}
\label{sec:conclusion}

This paper explores the development of generic TTSes for Indian languages in a unified end-to-end framework. It is observed that generic voices are quite robust and capable of capturing a wide variety of phonotactics. Incorporating speaker embedding in TTS provides flexibility in terms of speaker selection. For unseen speakers/languages, with as little as 7 minutes of adaptation data, generic voices are re-trained very quickly. These results are encouraging as they suggest the possibility of rapid adaption to a new language with minimal effort, especially useful for a low-resource scenario. Interactions between languages for applications such as code-switching can be further studied in the context of this work.
 

\vspace{-0.2cm}
\section{Acknowledgements}

The authors would like to thank the Ministry of Electronics and Information Technology (MeitY), Government of India, for funding the project, ``Text to Speech Generation with chosen accent and noise profile for Aerospace and Industrial domains'' (CSE1819172MIMPHEMA).

\bibliographystyle{IEEEtran}

\bibliography{Interspeech2020_v4}

\end{document}